\documentclass[10pt,journal]{IEEEtran}
\IEEEoverridecommandlockouts

\usepackage{graphicx}
\usepackage{amsthm,amsfonts,amssymb}
\usepackage[tbtags]{amsmath}
\usepackage{cite}
\usepackage{bm}
\usepackage{bbm}
\usepackage{url}
\usepackage{array}
\usepackage{color,soul}
\usepackage{multirow}
\usepackage{booktabs}
\usepackage[table,xcdraw]{xcolor}
\usepackage{enumitem}
\usepackage{subcaption}

\usepackage[english]{babel}
\usepackage{algorithm}
\usepackage[noend]{algpseudocode}
\usepackage{setspace}

\usepackage{comment}

\theoremstyle{plain}

\usepackage[english]{babel}
\usepackage{algorithm}
\usepackage[noend]{algpseudocode}

\allowdisplaybreaks[4]

\begin{document}

\title{From Link Diversity to Cross-Band Feedback Collaboration:\\ A New Perspective on Hybrid Optical-RF Systems}

\author{Menghan Li, Yulin Shao, Runxin Zhang, Lu Lu
\thanks{M. Li and L. Lu are with the Key Laboratory of Space Utilization, Technology and Engineering Center for Space Utilization, Chinese Academy of Sciences, and the University of Chinese Academy of Sciences, China.

Y. Shao and R. Zhang are with the Department of Electrical and Electronic Engineering, University of Hong Kong, Hong Kong S.A.R. 

For correspondence: ylshao@hku.hk.
}}



\maketitle

\begin{abstract}
We suggest a re-examination of the conventional view that hybrid optical-radio frequency (O-RF) systems are primarily diversity-driven networks that switch between RF and optical links for robustness. Instead, we uncover a new architectural opportunity: repurposing the optical downlink to enable real-time feedback channel coding over the RF uplink, where structured decoder feedback is delivered from the access point to guide the transmitter's coding strategy. This insight marks a conceptual paradigm shift from passive link diversity to active cross-band collaboration, where the wideband, interference-free optical wireless communication (OWC) is no longer merely a downlink backup but a functional enabler of uplink reliability.
To realize this vision, we propose a novel architecture, O-RF with Cross-Band Feedback (O-RF-CBF), that exploits the optical downlink feedback to facilitate adaptive RF uplink coding. Numerical results reveal that O-RF-CBF achieves significant uplink throughput gains over traditional O-RF systems. Our findings highlight that inter-band synergy, not redundancy, is the key to unlocking the full potential of hybrid wireless networks.
\end{abstract}

\begin{IEEEkeywords}
Hybrid optical-RF communication, optical communication, feedback-aided coding, resource allocation.
\end{IEEEkeywords}

\section{Introduction}
\begin{figure}[t]
  \centering
  \includegraphics[width=0.98\columnwidth]{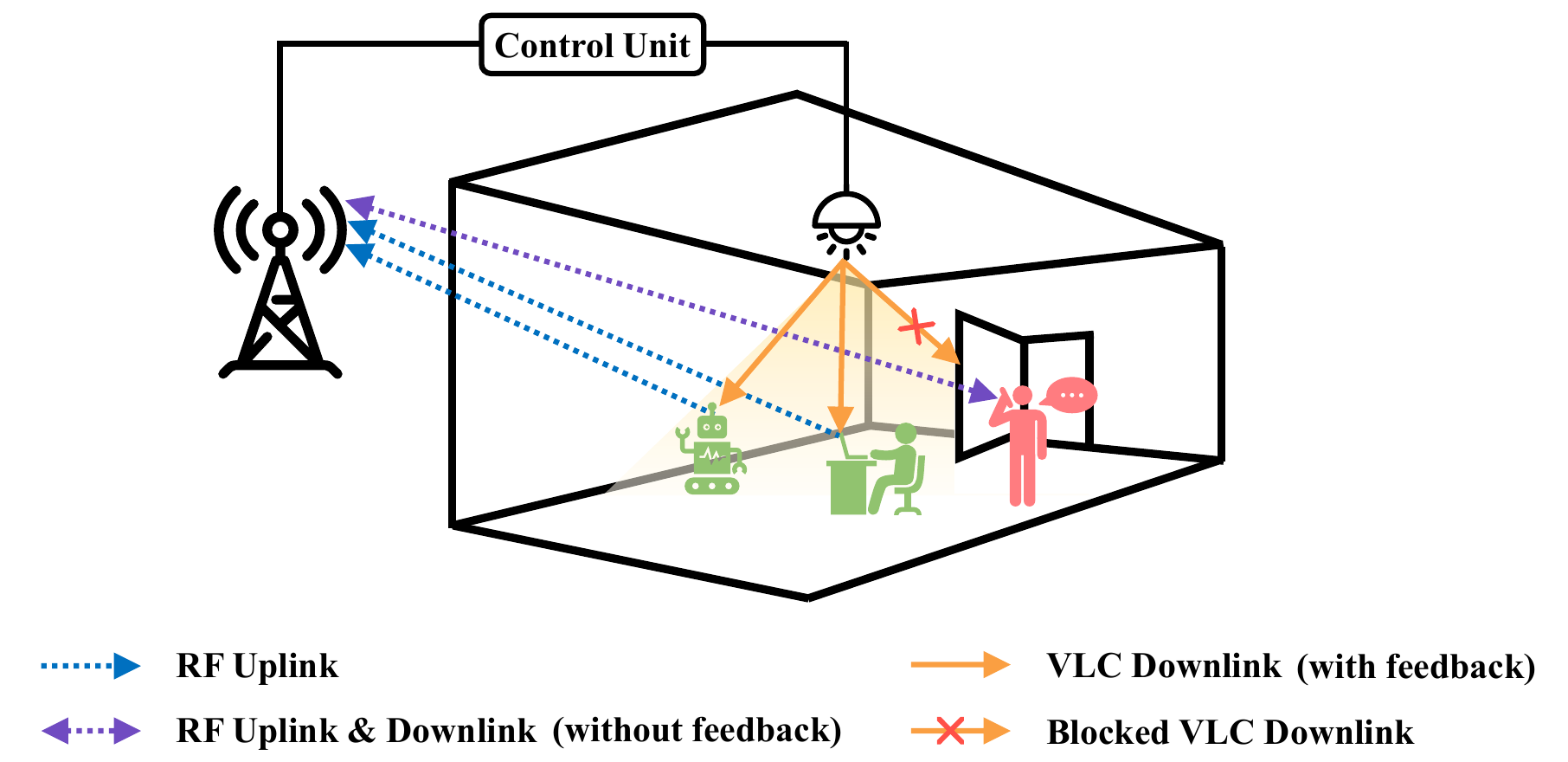}
  \caption{A typical hybrid O-RF communication deployment.}
\label{fig:sys}
\end{figure}

Hybrid optical-radio frequency (O-RF) communication systems have emerged as a promising solution for future wireless networks that demand high capacity, energy efficiency, and resilience\cite{chowdhury2020optical}. By integrating optical wireless communication (OWC), particularly visible light communication (VLC), with radio frequency (RF) transmission, these systems exploit the complementary strengths of both domains: the abundant, interference-free spectrum of VLC and the omnidirectional, mobility-friendly nature of RF\cite{tang2024channel,zhang2024optical}. A typical hybrid deployment, as illustrated in Fig.~\ref{fig:sys}, utilizes VLC for downlink communication (leveraging existing lighting infrastructure) and RF for uplink, aligning with the asymmetric traffic patterns commonly found in Internet of Things (IoT) applications.

Prior research on O-RF systems primarily interprets the optical and RF links as parallel downlink transmission paths, focusing on ``link-level diversity'' to enhance robustness and throughput\cite{chowdhury2020optical,rallis2023energy}. Common strategies include switching between VLC and RF under poor channel conditions (fallback)\cite{bao2018vertical}, distributing traffic across both links (load balancing)\cite{wang2023hybrid}, or aggregating bandwidth for higher data rates (aggregation)\cite{Aboagye2023energy}. While these designs offer flexibility and reliability, they fundamentally treat the two links as interchangeable data carriers without assigning differentiated functional roles or enabling inter-link cooperation.

In this paper, we propose a new perspective on hybrid O-RF systems:  we reveal that {\it the wideband, interference-free OWC is not only well-suited for downlink data transmission to relieve RF spectrum congestion, but also naturally positioned to support real-time feedback to directly aid uplink channel coding.} This insight enables a novel form of ``cross-band feedback collaboration'', wherein the optical downlink is leveraged to deliver timely feedback information, such as the real-time decoding status of the access point (AP), to the IoT device, allowing the device to modify its encoding schemes for subsequent transmissions. In turn, this feedback empowers the device to adapt its RF uplink encoding strategy on the fly, dynamically adjusting its transmission based on receiver-side conditions\cite{shao2025deep}. Such feedback channel coding, inspired by schemes like Schalkwijk-Kailath\cite{schalkwijk1966coding} and DL-based coding schemes\cite{kim2020deepcode,shao2024theory}, can substantially improve the reliability and rate of RF uplink communication.

The main contributions of this paper are as follows:
\begin{itemize}[leftmargin=0.4cm]
    \item We propose a novel O-RF architecture with cross-band feedback coding (O-RF-CBF), where the optical downlink is utilized not only for high-rate data transmission but also as a channel for delivering real-time feedback information, which enables dynamic adaptation of uplink channel coding strategy, thereby enhancing RF uplink reliability. This design departs from traditional link-diversity-based O-RF systems and introduces a new form of functional inter-band cooperation, shifting the architectural paradigm from resource redundancy to synergistic integration. 
    \item We develop a tractable analytical framework to quantify the performance of O-RF-CBF under practical system constraints, including probabilistic optical blockage and feedback-induced payload cost. We formulate a sum-log throughput maximization problem that jointly optimizes the allocation of channel symbols and information bits per device. The solution approach leverages alternating optimization, structural analysis of the PER functions, and successive convex approximation to efficiently handle the non-convexity and integer constraints.
    \item Through extensive numerical results, we show that O-RF-CBF achieves significant uplink throughput gains over conventional O-RF systems, especially under asymmetric channel conditions and moderate feedback cost. Our analysis also characterizes the impact of optical blockage, number of users, and feedback overhead, offering actionable insights for the design of future hybrid communication systems.
\end{itemize}

\section{System Model}
We consider a hybrid O-RF communication system consisting of an AP and $L$ IoT devices, as illustrated in Fig.~\ref{fig:sys}.
Each device maintains two separate communication links with the AP: an RF link and an optical link. 
The RF link supports both uplink and downlink communication, but its resource is limited due to spectrum congestion. In contrast, the optical link can only support downlink transmission, leveraging existing lighting infrastructure.

Compared with the RF link, the optical downlink provides abundant bandwidth and strong line-of-sight (LOS) signal concentration. As such, when unobstructed, the optical link can be considered ideal. However, its vulnerability to blockage introduces binary link behavior. We denote by $p_\ell \in [0,1]$ the probability that the downlink from the AP to device $\ell$ is blocked.

In this work, we focus on the uplink transmission and evaluate the system over a total communication duration of $T$ seconds. Suppose there are $N$ RF channel uses available during $T$. Each device $\ell \in {1,\ldots,L}$ is allocated $N_\ell$ uplink channel symbols to transmit $K_\ell$ information bits to the AP.

The signal received by the AP from device $\ell$ on the $n$-th allocated channel is modeled as
\begin{equation}\label{eq:ULsignal}
y_{\ell,n} = \sqrt{P_{\ell}} h_{\ell} s_{\ell,n} + w_{\ell,n},
\end{equation}
where $s_{\ell,n}$ is the transmitted symbol normalized to unit average power, $P_\ell$ is the transmit power, $h_\ell$ is the quasi-static Rayleigh fading gain, and $w_{\ell,n} \sim \mathcal{N}(0, \sigma^2)$ denotes additive white Gaussian noise (AWGN).
The resulting uplink signal-to-noise ratio (SNR) is given by
\begin{equation}
\eta_\ell = \frac{|h_\ell|^2 P_\ell}{\sigma^2}.
\end{equation}

In conventional O-RF systems employing forward (non-feedback) channel coding, the uplink packet error rate (PER) can be obtained by the finite blocklength coding theorem \cite{polyanskiy2010channel}:
\begin{equation}\label{PER_f}
\varepsilon_\ell = Q\left( \frac{N_\ell \log_2(1 + \eta_\ell) - 2K_\ell}{\log_2 e \cdot \sqrt{ \frac{2 N_\ell \eta_\ell(\eta_\ell + 2)}{(\eta_\ell + 1)^2} } } \right),
\end{equation}
where $Q(\cdot)$ is the tail probability of the standard Gaussian distribution. The corresponding throughput of device $\ell$ is
\begin{equation}
r^{(f)}_\ell = \frac{K_\ell (1 - \varepsilon_\ell)}{T}.
\end{equation}

In the proposed O-RF-CBF framework, we utilize the optical downlink, when unblocked, to deliver real-time feedback from the AP to IoT devices. This enables feedback-aided channel coding over the RF uplink, allowing devices to dynamically adjust their coding strategies based on the AP's decoding status. To quantify the performance improvement, this paper assumes the SK feedback coding scheme, under which the PER can be characterized by \cite{schalkwijk1966coding}
\begin{equation}\label{PER_b}
\epsilon_\ell = \frac{2^{K_\ell} - 1}{2^{K_\ell - 1}} Q\left( \sqrt{ \frac{3}{4^{K_\ell} - 1} \left( \eta_\ell + \frac{N_\ell - 1}{N_\ell} \right)^{N_\ell} } \right).
\end{equation}
In the feedback mode, a fraction of the transmission resource is reserved for feedback signaling, reducing the effective payload. Denoting by $\zeta \in (0,1)$ the normalized feedback-induced payload cost, the throughput in the feedback mode can be written as
\begin{equation}\label{rb}
r^{(b)}_{\ell} = \frac{K_\ell (1 - \epsilon_\ell)}{T}(1-\zeta).
\end{equation}

Accounting for the stochastic blockage of the optical link, the average uplink throughput of device $\ell$ under  O-RF-CBF becomes
\begin{equation}
r_{\ell} = p_\ell r^{(f)}_{\ell} + (1 - p_\ell) r^{(b)}_{\ell}.
\end{equation}
Note that the conventional O-RF system is a special case of O-RF-CBF by setting $p_\ell = 1$ for all devices. Therefore, in the remainder of this paper, we focus on analyzing $r_{\ell}$.

To quantify the performance gains of O-RF-CBF, we adopt the sum-log throughput as a utility metric, defined as $\sum_{\ell=1}^{L} \ln r_\ell$, which promotes proportional fairness among devices \cite{lau2005proportional}. Accordingly, the optimal sum-log throughput achievable by O-RF-CBF can be obtained by solving the following problem:
\begin{subequations}\label{opt:main}
\begin{align}
\max_{{K_\ell, N_\ell}} &~ \sum_{\ell=1}^{L} \ln r_\ell, \label{obj} \\
\text{s.t.}~~ & \sum_{\ell=1}^{L} N_\ell = N, \label{N_total} \\
& K_\ell, N_\ell \in \mathbb{Z}^+,~ \forall \ell.
\end{align}
\end{subequations}

\section{Optimal Sum-log Throughput}
In this section, we solve the optimization problem \eqref{opt:main} to derive the optimal sum-log throughput achievable by the proposed O-RF-CBF framework.
The problem in \eqref{opt:main} poses two main challenges: a non-convex objective function and integer constraints on both $K_\ell$ and $N_\ell$. To address this, we adopt a two-step approach: 
i) relax the integer constraints to obtain a tractable continuous approximation; and 
ii) apply alternating optimization to iteratively update the coupled variables $\{K_\ell\}$ and $\{N_\ell\}$. The resulting continuous solution is then rounded to satisfy the original integer constraints.

We begin by relaxing the integer constraints and focusing on optimizing $\{K_\ell\}$ while holding $\{N_\ell\}$ fixed. The resulting subproblem becomes
\begin{subequations}\label{opt:K}
\begin{align}
\max_{{K_\ell}} &~ \sum_{\ell=1}^{L} \ln r_\ell, \\
\text{s.t.}~~ & K_\ell \geq 1,~ \forall \ell. 
\end{align}
\end{subequations}

Problem \eqref{opt:K} remains non-convex due to the presence of the $Q$-function in both the forward- and feedback-mode PER $\varepsilon_\ell$ and $\epsilon_\ell$. Nonetheless, we can efficiently identify the optimal solution by examining the structure of the objective function.
In particular, we observe that the optimal value of each $K_\ell$ can be obtained by locating the zero of the first-order derivative of the corresponding throughput $r_\ell$. This observation is enabled by two key properties:
\begin{itemize}[leftmargin=0.4cm]
    \item There are no coupling constraints across devices in the subproblem \eqref{opt:K}, allowing each $K_\ell$ to be optimized independently.
    \item The logarithmic operation in the objective does not alter the location of local extrema; maximizing $\ln r_{\ell}$ is equivalent to maximizing $r_\ell$ itself with respect to $K_\ell$.
\end{itemize}

With these properties, the original subproblem \eqref{opt:K} is equivalent to solving the following per-device optimization:
\begin{subequations}\label{opt:K_l}
\begin{align}
&\sum_{\ell=1}^{L} ~ \max_{K_\ell} r_\ell, \\
&~\text{s.t.} ~~K_\ell \geq 1,~ \forall \ell. 
\end{align}
\end{subequations}

To determine the optimal payload size $K_\ell$, we analyze the derivative of $r_\ell$ with respect to $K_\ell$, yielding
\begin{eqnarray}\label{eq:dr}
\frac{\partial r_\ell}{\partial K_\ell} \hspace{-0.3cm}&=&\hspace{-0.3cm} (1-p_\ell)\left(1-2Q\left(\sqrt{\frac{A}{4^{K_\ell}}}\right)\right)+p_\ell Q\left(\frac{2K_\ell-C}{B}\right) \nonumber\\
&&\hspace{-1.35cm} -K_\ell\left(\left(1\!-\!p_\ell\right)\sqrt{\frac{2A}{\pi}}\ln{2}\cdot\frac{e^{-\frac{A}{2\cdot4^{K_\ell}}}}{2^{K_\ell}}\!+\!\frac{p_\ell}{B}\sqrt{\frac{2}{\pi}}\cdot e^{-\frac{(2K_\ell\!-\!C)^2}{2B^2}}\right) \nonumber\\
&&\hspace{-1.35cm}\approx\hspace{-0.1cm} (1-p_\ell)\left(1-e^{-\frac{A}{2^{2K_\ell+1}}}\right)+p_\ell Q\left(\frac{2K_\ell-C}{B}\right) \nonumber\\
&&\hspace{-1cm}-\frac{p_\ell K_\ell}{B}\cdot e^{-\frac{(2K_\ell-C)^2}{2B^2}} \nonumber\\
&&\hspace{-1.35cm}\overset{(a)}{=} (1-p_\ell)\left(1-e^{-\frac{A}{2^{Bz+C+1}}}\right)+p_\ell Q(z)-\frac{p_\ell}{2}(z+\lambda)e^{-\frac{z^2}{2}} \nonumber\\
&&\hspace{-1.35cm} \triangleq g(z),
\end{eqnarray}
where $A\triangleq 3\left(\eta_\ell+\frac{N_\ell-1}{N_\ell}\right)^{N_\ell}$, $B\triangleq \log_2 e\cdot\frac{\sqrt{2N_\ell\eta_\ell(\eta_\ell+2)}}{\eta_\ell+1}$, $C\triangleq N_\ell\log_2 (1+\eta_\ell)$, and $(a)$ follows by defining $z\triangleq \frac{2K_\ell-C}{B}$ and $\lambda\triangleq \frac{C}{B}$. Taking the derivative of $g(z)$, we obtain:
\begin{eqnarray}
g'(z) \hspace{-0.3cm}&=&\hspace{-0.3cm} -A\ln{2}\cdot (1-p_\ell)e^{-\frac{A}{2^{Bz+C+1}}}\cdot2^{-Bz-C} \nonumber\\
&&\hspace{0.1cm} -\frac{p_\ell}{2}e^{-\frac{z^2}{2}}\left(-z(z+\lambda)+\sqrt{{2}/{\pi}}+1\right) \\
&\approx&\hspace{-0.3cm}\frac{p_\ell}{2}e^{-\frac{z^2}{2}}\left(z^2+\lambda z-\sqrt{{2}/{\pi}}-1\right) \overset{(b)}{=} \frac{p_\ell}{2}e^{-\frac{z^2}{2}}\cdot h(z), \nonumber
\end{eqnarray}
where $(b)$ follows by defining $h(z)\triangleq z^2+\lambda z-\sqrt{\frac{2}{\pi}}-1$.

Setting $g'(z)=0$ is thus equivalent to solving $h(z)=0$, which yields $z=\frac{-\lambda\pm\sqrt{\lambda^2+4(\sqrt{2/\pi+1})}}{2}$.
Among the two roots, only the positive root $z^+=\frac{-\lambda+\sqrt{\lambda^2+4(\sqrt{2/\pi+1})}}{2}$ lies in the feasible domain of interest.
From the sign of $h(z)$, it follows that: $g'(z)<0$ when $z<z^+$; $g'(z)>0$ when $z>z^+$.

Therefore, $g(z)$ is strictly decreasing then strictly increasing, hence unimodal. Next, we analyze the limiting behavior of the derivative. We have
\begin{eqnarray*}
\hspace{-0.65cm} && \lim_{K\to0}\frac{\partial r_\ell}{\partial K_\ell}\!=\!(1\!-\!p_\ell)\left(1\!-\!e^{-\frac{A}{2^{2K_\ell+1}}}\right)\!+\!p_\ell Q\left(\frac{2K_\ell\!-\!C}{B}\right)\!>\!0, \\
\hspace{-0.6cm} && \lim_{K\to \infty}-\frac{p_\ell K_\ell}{B}e^{-\frac{(2K_\ell-C)^2}{2B^2}}<0.
\end{eqnarray*}

As a result, $r_{\ell}(K_\ell)$ is unimodal and has a unique global maximizer. This enables efficient one-dimensional search (e.g., via Newton-Raphson or bisection) to determine the optimal $K_\ell^*$ given any fixed $N_{\ell}$, where
\begin{equation}\label{optimalK}
K_\ell^* = \left\{ K_\ell \middle| \frac{d r_\ell}{d K_\ell} = 0 \right\}, \quad \forall \ell.
\end{equation}

In the next step, we consider the optimization of the resource allocation variables $\{N_\ell\}$ while fixing $\{K_\ell\}$ at the values obtained in the previous step. Under this setting, the sum-log throughput maximization problem reduces to
\begin{subequations}\label{opt:N}
\begin{align}
\max_{{N_\ell}} &~ \sum_{\ell=1}^{L} \ln \widetilde{r}_\ell, \\
\text{s.t.} ~~&\sum_{\ell=1}^{L} N_\ell = N, \label{eq:14b} \\ 
& \label{eq:14c} N_\ell \geq 1,~ \forall \ell. 
\end{align}
\end{subequations}
Similarly, the $Q$-function in the objective function makes problem \eqref{opt:N} non-convex. To enable tractable analysis, we adopt an exponential approximation of the Gaussian $Q$-function:
\begin{equation}\label{eq:approx_Q}
Q(x)\approx\left\{
\begin{array}{l}
   \frac{1}{2}e^{-\frac{x^2}{2}}, \hspace{1.4cm}x\geq0, \\   
   1-\frac{1}{2}e^{-\frac{x^2}{2}}, \hspace{0.8cm}x<0.
\end{array}
\right.
\end{equation}

Using this approximation, we simplify the expressions for the PERs. Starting with the feedback-aided mode, we observe that its argument is always non-negative. Thus, we use the $x \geq 0$ branch of \eqref{eq:approx_Q} to approximate \eqref{PER_b} as
\begin{equation}\label{eq:approx_PER_b}
\epsilon_\ell\approx
   \frac{1}{2}\exp\left(-\frac{3\left(\eta_\ell+\frac{N_\ell-1}{N_\ell}\right)^{N_\ell}}{2\left(4^{K_\ell}-1\right)}\right).
\end{equation}
For the non-feedback case in \eqref{PER_f}, the choice of approximation depends on the sign of the expression inside $Q(\cdot)$, which is determined by the comparison between $N_\ell$ and $\frac{2K_\ell}{\log_2(1 + \eta_\ell)} $. Specifically:
\begin{itemize}[leftmargin=0.4cm]
    \item When $N_\ell \leq \frac{2K_\ell}{\log_2(1 + \eta_\ell)}$, we apply the $x \geq 0$ branch:
    \begin{equation*}
\varepsilon_\ell\approx
   \frac{1}{2}\exp\left(-\frac{\left[2K_\ell-N_\ell\log_2 (\eta_\ell+1)\right]^2(\eta_\ell+1)^2}{4(\log_2 e)^2N_\ell\eta_\ell(\eta_\ell^2+2)}\right).
\end{equation*}
   \item When $N_\ell > \frac{2K_\ell}{\log_2(1 + \eta_\ell)}$, we apply the $x < 0$ branch:
\begin{equation*}
\varepsilon_\ell\approx
   1-\frac{1}{2}\exp\left(-\frac{\left[2K_\ell-N_\ell\log_2 (\eta_\ell+1)\right]^2(\eta_\ell+1)^2}{4(\log_2 e)^2N_\ell\eta_\ell(\eta_\ell^2+2)}\right).
\end{equation*}
\end{itemize}

Although the above two expressions differ in form depending on the coding rate region, they share a consistent mathematical structure and allow the objective function to be expressed in closed form. For clarity and without loss of generality, we focus the subsequent analysis on the case $N_\ell \leq \frac{2K_\ell}{\log_2(1 + \eta_\ell)}$. The analysis for the alternative case proceeds similarly.

Let $\widetilde{r}_\ell$ denote the throughput $r_\ell$ after substituting $Q(\cdot)$ with its approximation. Despite this simplification, the objective $\sum_{\ell=1}^{L} \ln \widetilde{r}_\ell$ remains non-convex. To address this issue, we introduce an auxiliary variable $\alpha_\ell$ and reformulate problem \eqref{opt:N} as
\begin{subequations}\label{opt:alpha}
\begin{align}
\max_{{\alpha_\ell, N_\ell}} &~ \sum_{\ell=1}^{L} \ln \alpha_\ell,\\
\text{s.t.}~~ & \alpha_\ell \leq \widetilde{r}_\ell,~ \forall \ell, \\
& N_\ell \leq \frac{2K_\ell}{\log_2 (1+\eta_\ell)},~ \forall \ell, \\
& (14b), (14c). \notag
\end{align}
\end{subequations}

Problem \eqref{opt:alpha} is still non-convex due to the nonlinear constraint involving $\widetilde{r}_\ell(N_\ell)$. To convexify it, we apply the successive convex approximation (SCA) technique. Specifically, we construct a first-order Taylor expansion of $\widetilde{r}_\ell$ around the point $(N_\ell^i, \widetilde{r}_\ell^i)$ obtained at iteration $i$, and add a regularization term to ensure the surrogate function is a lower bound of   $\widetilde{r}_\ell$, yielding
\begin{equation}\label{Taylor}
\widetilde{r}_\ell\leq\widetilde{r}_\ell^i+\nabla\widetilde{r}_\ell^i\left(N_\ell-N_\ell^i\right)+\frac{\rho}{2}\left(N_\ell-N_\ell^i\right)^2,
\end{equation}
where $\rho\leq0$ is a regularization parameter controlling curvature. Denote the second-order Taylor expansion coefficient of $\widetilde{r}_\ell$ as $\rho_0$, and then $\rho=\min(\rho_0,0)$.

Substituting this surrogate into \eqref{opt:alpha}, we obtain a convexified version of the subproblem:
\begin{subequations}\label{opt:p1}
\begin{align}
\text{(P1)}:~~ &~ \max_{{\alpha_\ell, N_\ell}} \sum_{\ell=1}^{L} \ln \alpha_\ell,\\
\text{s.t.}~~  \alpha_\ell &\leq \widetilde{r}_\ell^i+\nabla\widetilde{r}_\ell^i\left(N_\ell-N_\ell^i\right) +\frac{\rho}{2}\left(N_\ell\!-\!N_\ell^i\right)^2,~ \forall \ell, \\
&\hspace{-0.4cm}  (14b), (14c), (17c). \notag
\end{align}
\end{subequations}
Problem (P1) can be solved using standard convex solvers at each iteration of the SCA algorithm. For the complementary case $N_\ell > \frac{2K_\ell}{\log_2 (1+\eta_\ell)}$, the transformation of problem \eqref{opt:N} follows an analogous procedure. Given the structural similarity, we omit the details for brevity and denote the resulting convex formulation as subproblem (P2).

Once convergence is achieved over the alternating updates of $\{K_\ell\}$ and $\{N_\ell\}$, we revisit the integer constraints originally relaxed during optimization. Since the problem involves $2L$ integer variables and the objective function is complicated, exhaustive enumeration through classic methods such as branch-and-bound or cutting plane becomes computationally infeasible. Instead, we adopt a lightweight combinatorial refinement strategy. Specifically, for each continuous solution pair $(K_\ell, N_\ell)$, we consider its upward and downward integer roundings, leading to $2^{2L}$ candidate combinations. Among these, we discard any combination that violates the total bandwidth constraint $\sum_{\ell=1}^{L} N_\ell = N$. For the remaining feasible combinations, we evaluate the original (non-approximated) objective in \eqref{obj} and select the one yielding the highest sum-log throughput as the final solution.

\section{Numerical results}
In this section, we evaluate the performance of the proposed O-RF-CBF system compared with the conventional O-RF system that lacks cross-band feedback. Unless otherwise specified, the total number of uplink RF channel symbols is set to $N = 256$; the blockage probability of each device's optical link $p_\ell\in[0,1]$; the normalized feedback cost is $\zeta = 0.05$; the RF links follow independent quasi-static Rayleigh fading; 

\begin{figure}[t]
  \centering
  \includegraphics[width=0.9\columnwidth]{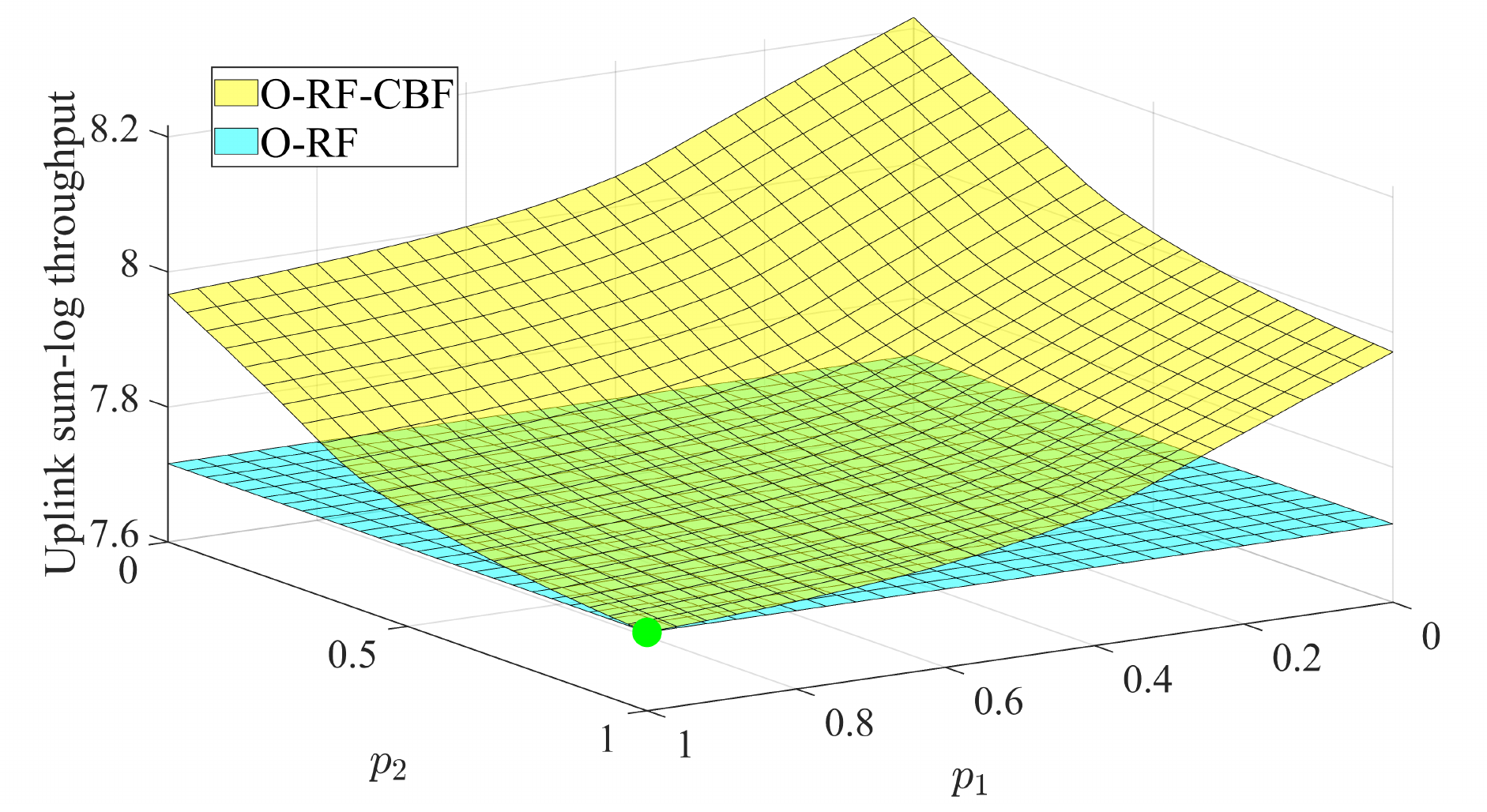}
  \caption{The uplink sum-log throughput versus the block probabilities $p_1$ and $p_2$ in a small-scale network with $L=2$.}
\label{fig:r-p}
\end{figure}

We begin with a small-scale network consisting of $L=2$ devices. Fig.~\ref{fig:r-p} illustrates the uplink sum-log throughput achieved by O-RF and O-RF-CBF as a function of the optical blockage probabilities $p_1$ and $p_2$. When $p_1 = p_2 = 1$, both optical links are fully blocked, rendering the VLC downlink unusable. In this case, the two systems yield identical throughput, as no feedback is available. However, as either $p_1$ or $p_2$ decreases, indicating more frequent availability of the optical downlink, the throughput of O-RF-CBF increases significantly, benefiting from the real-time feedback enabled by cross-band collaboration. By contrast, the performance of the conventional O-RF system remains constant across all $p_\ell$ values, as it lacks any mechanism to exploit improved optical links.

\begin{figure}[t]
  \centering
  \includegraphics[width=0.9\columnwidth]{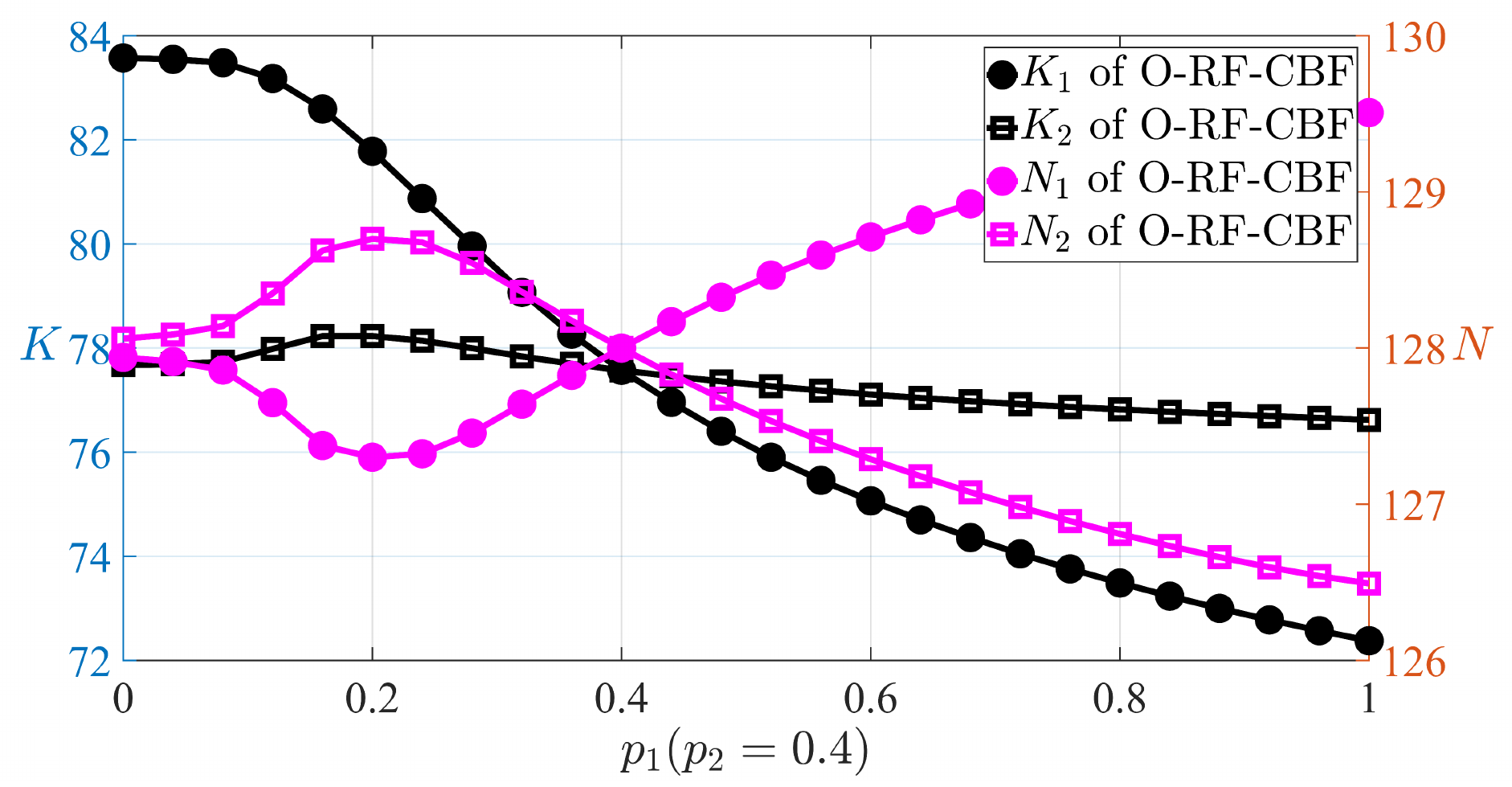}
  \caption{The optimized allocated transmitted bits $K_1$, $K_2$ and channel symbols $N_1$, $N_2$ (before rounding) versus $p_1$, where $p_2=0.4$.}
\label{fig:iflofp}
\end{figure}

To better understand the resource allocation dynamics, we examine a cross-section of Fig.~\ref{fig:r-p} by fixing $p_2 = 0.4$ and varying $p_1$. Fig.~\ref{fig:iflofp} shows the optimized values of the transmitted bits $K_1, K_2$ and allocated channel symbols $N_1, N_2$ for the two devices. As $p_1$ increases, device 1 becomes less capable of receiving feedback, prompting the system to allocate more RF symbols ($N_1$) but fewer bits ($K_1$) to mitigate higher error probability. Due to the fixed total symbol budget, this reallocation reduces $N_2$ for device 2, although the impact on $K_2$ is more modest thanks to its stable optical channel. When $p_1 = p_2 = 0.4$, the link conditions are symmetric, leading to a balanced allocation. Interestingly, for small $p_1<0.24$, a slight increase in $p_1$ initially results in fewer symbols allocated to device 1. This counterintuitive trend stems from the nonlinearity of the sum-log throughput function: in this regime, diverting more resources to device 2 yields a higher marginal utility, illustrating the delicate interplay between channel quality and fairness constraints.

\begin{figure}[t]
  \centering
  \includegraphics[width=0.9\columnwidth]{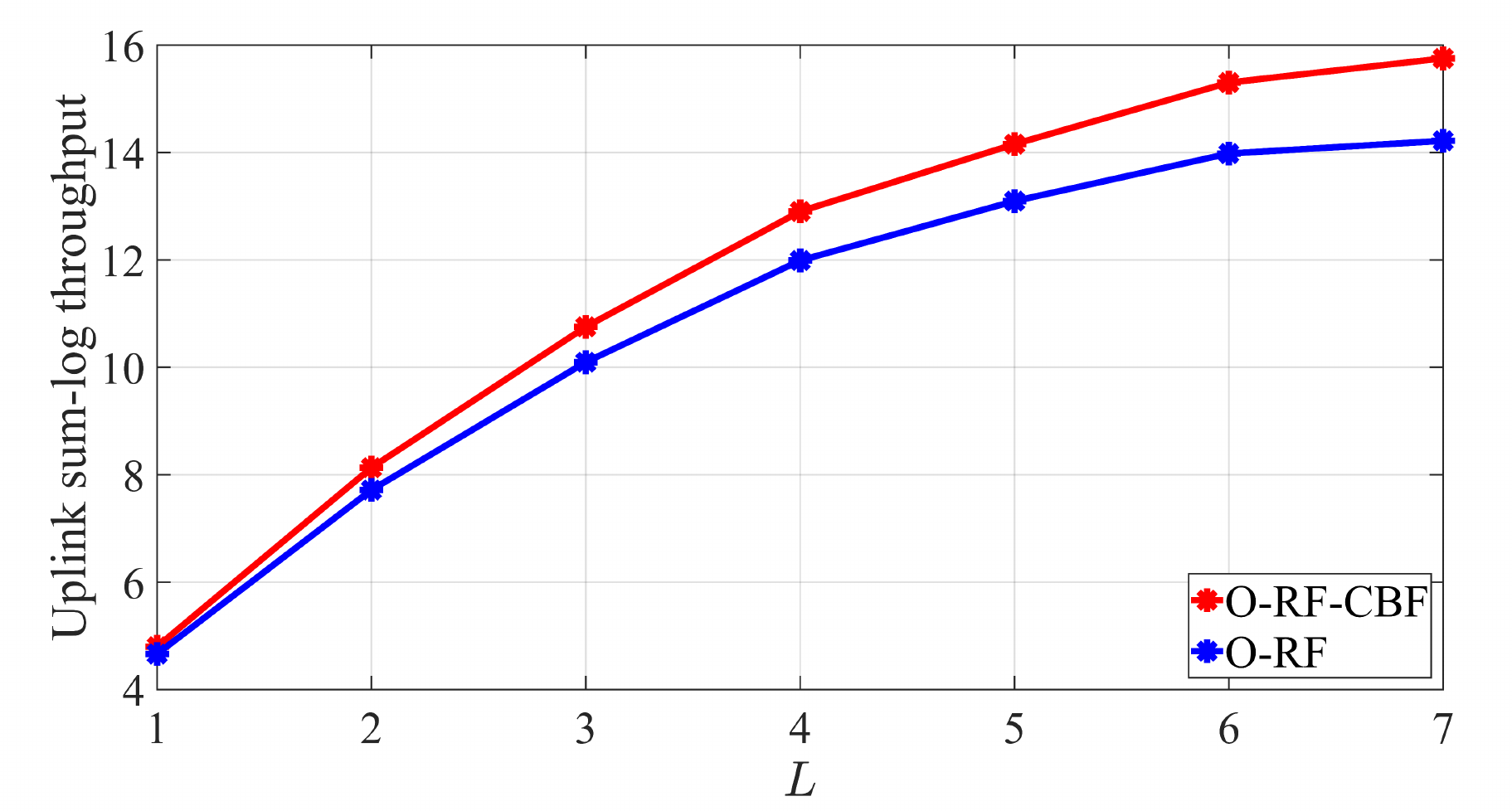}
  \caption{Impact of the number of devices $L$ on the uplink sum-log throughput, where $\{p_\ell\}=0.1$, $\forall \ell$.}
\label{fig:r-L}
\end{figure}

Next, we evaluate the scalability of the O-RF-CBF architecture in larger networks. Fig.~\ref{fig:r-L} presents the uplink sum-log throughput as the number of devices $L$ increases, assuming all devices have identical blockage probabilities $p_\ell = 0.1$. It can be observed that the sum-log throughput increases with growing $L$ and the proposed O-RF-CBF system consistently outperforms the baseline O-RF design.
Moreover, the performance gap widens with larger $L$, as more devices can benefit from the cross-band feedback provided by the optical downlink. This highlights the inherent scalability of O-RF-CBF and its potential in dense IoT deployments.

\begin{figure}[t]
  \centering
  \includegraphics[width=0.9\columnwidth]{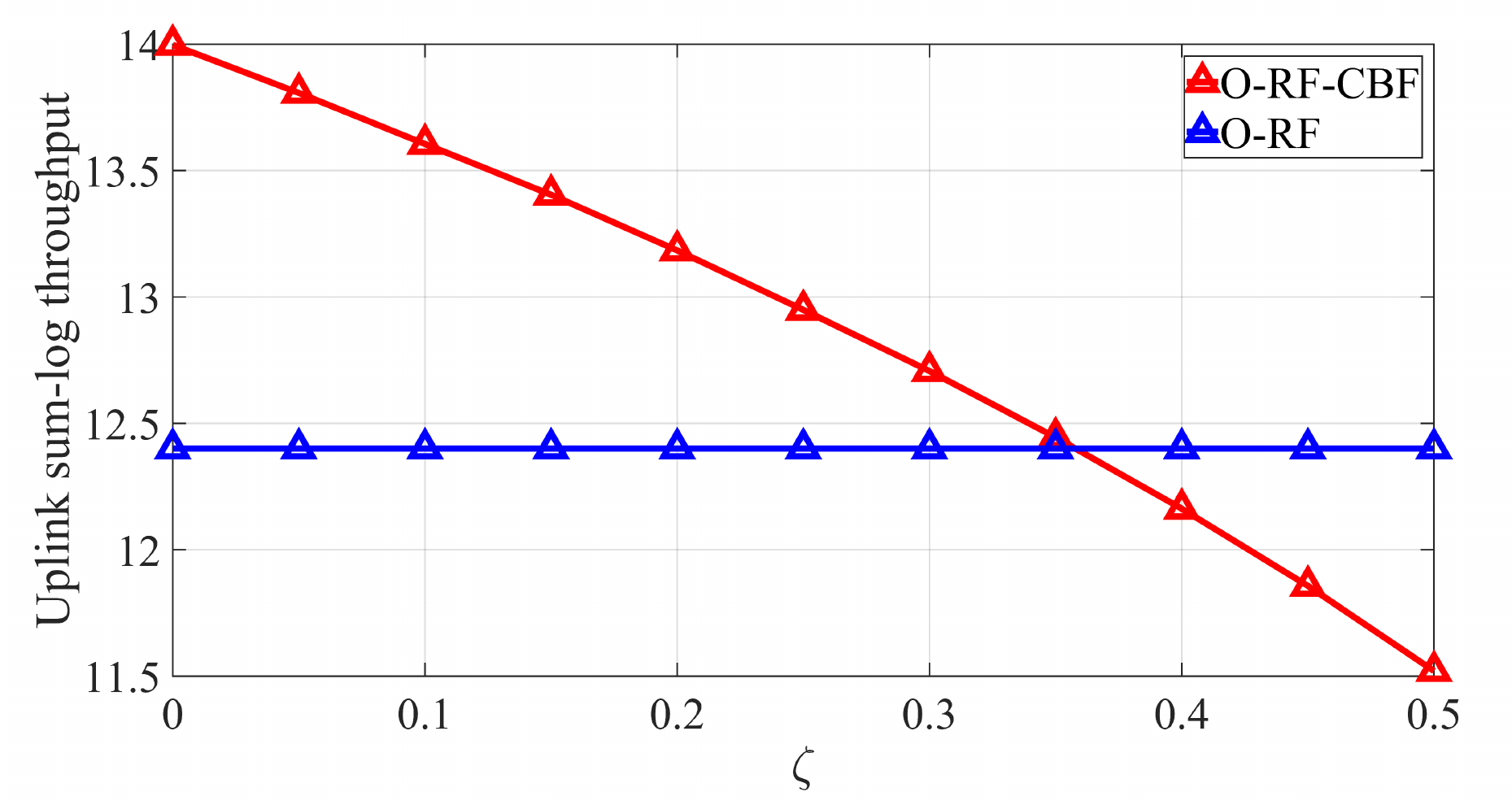}
  \caption{Impact of the normalized feedback cost $\zeta$ on the uplink sum-log throughput, where $L=4$ and $\{p_\ell\}=0.1$, $\forall \ell$.}
\label{fig:r-zeta}
\end{figure}

It is important to note that the gains of O-RF-CBF does not come for free. The downlink feedback introduced unavoidable resource consumption, reflected by the normalized payload cost $\zeta$. Considering $L=4$, Fig.~\ref{fig:r-zeta} illustrates the uplink sum-log throughput with the increase in $\zeta$. As shown, the sum-log throughput of O-RF-CBF decreases monotonically with increasing $\zeta$. On the other hand, O-RF does not be infected by the feedback cost and the throughput remains constant. When $\zeta>0.35$, the cost surpass the gains, and the cross-band collaboration is no longer worthwhile.

While the proposed O-RF-CBF architecture demonstrates clear advantages over traditional O-RF systems, it is important to evaluate the impact of its feedback overhead. Fig.~\ref{fig:r-zeta} investigates how the uplink sum-log throughput varies with the normalized feedback cost $\zeta$. As shown, the throughput of O-RF-CBF decreases gradually with increasing $\zeta$, due to the reduced effective payload. In contrast, the baseline O-RF system maintains constant performance.
Nevertheless, even under moderate feedback costs (e.g., $\zeta \leq 0.1$), O-RF-CBF continues to offer substantial throughput gains over O-RF. This highlights the efficiency of leveraging the optical downlink for feedback, which consumes only a negligible portion of optical resources while enabling significant uplink enhancements. When $\zeta$ becomes excessively large ($\zeta > 0.35$), the marginal cost may outweigh the benefit.

\section{Conclusion}
This work reexamines hybrid optical-RF systems through the lens of functional cooperation, rather than redundancy. Our results demonstrate that enabling real-time feedback via the optical downlink fundamentally improves uplink reliability and resource utilization. More importantly, this finding shifts the design principle of hybrid systems: inter-band feedback is not an auxiliary feature, but a central architectural enabler.

\bibliographystyle{IEEEtran}
\bibliography{ref.bib}

\begin{thebibliography}{10}
\providecommand{\url}[1]{#1}
\csname url@samestyle\endcsname
\providecommand{\newblock}{\relax}
\providecommand{\bibinfo}[2]{#2}
\providecommand{\BIBentrySTDinterwordspacing}{\spaceskip=0pt\relax}
\providecommand{\BIBentryALTinterwordstretchfactor}{4}
\providecommand{\BIBentryALTinterwordspacing}{\spaceskip=\fontdimen2\font plus
\BIBentryALTinterwordstretchfactor\fontdimen3\font minus \fontdimen4\font\relax}
\providecommand{\BIBforeignlanguage}[2]{{%
\expandafter\ifx\csname l@#1\endcsname\relax
\typeout{** WARNING: IEEEtran.bst: No hyphenation pattern has been}%
\typeout{** loaded for the language `#1'. Using the pattern for}%
\typeout{** the default language instead.}%
\else
\language=\csname l@#1\endcsname
\fi
#2}}
\providecommand{\BIBdecl}{\relax}
\BIBdecl

\bibitem{chowdhury2020optical}
M.~Z. Chowdhury, M.~K. Hasan, M.~Shahjalal, M.~T. Hossan, and Y.~M. Jang, ``Optical wireless hybrid networks: Trends, opportunities, challenges, and research directions,'' \emph{IEEE Communications Surveys \& Tutorials}, vol.~22, no.~2, pp. 930--966, 2020.

\bibitem{tang2024channel}
P.~Tang, Y.~Yin, Y.~Tong, S.~Liu, L.~Li, T.~Jiang, Q.~Wang, and M.~Chen, ``Channel characterization and modeling for {VLC}-{I}o{E} applications in 6{G}: A survey,'' \emph{IEEE Internet of Things Journal}, 2024.

\bibitem{zhang2024optical}
R.~Zhang, Y.~Shao, M.~Li, L.~Lu, and Y.~C. Eldar, ``Optical integrated sensing and communication with light-emitting diode,'' \emph{IEEE Internet of Things Journal}, vol.~12, no.~9, pp. 12\,896--12\,911, 2025.

\bibitem{rallis2023energy}
K.~G. Rallis, V.~K. Papanikolaou, P.~D. Diamantoulakis, S.~A. Tegos, A.~A. Dowhuszko, M.-A. Khalighi, and G.~K. Karagiannidis, ``Energy efficient cooperative communications in aggregated {VLC/RF} networks with {NOMA},'' \emph{IEEE Trans. Comm.}, vol.~71, no.~9, pp. 5408--5419, 2023.

\bibitem{bao2018vertical}
X.~Bao, W.~Adjardjah, A.~Okine, W.~Zhang, and N.~Bao, ``Vertical handover scheme for enhancing the {Q}o{E} in {VLC} heterogeneous networks,'' in \emph{IEEE/CIC ICCC}, 2018.

\bibitem{wang2023hybrid}
F.~Wang, F.~Yang, C.~Pan, J.~Song, and Z.~Han, ``Hybrid {VLC-RF} systems with multi-users for achievable rate and energy efficiency maximization,'' \emph{IEEE Trans. Wireless Comm.}, vol.~22, no.~9, pp. 6157--6170, 2023.

\bibitem{Aboagye2023energy}
S.~Aboagye, T.~M.~N. Ngatched, O.~A. Dobre, and H.~V. Poor, ``Energy-efficient resource allocation for aggregated {RF/VLC} systems,'' \emph{IEEE Trans. Wireless Comm.}, vol.~22, no.~10, pp. 6624--6640, 2023.

\bibitem{shao2025deep}
Y.~Shao, ``{DEEP}-{I}o{T}: Downlink-enhanced efficient-power {I}nternet of {T}hings,'' \emph{IEEE Trans. Wireless Comm.}, 2025.

\bibitem{schalkwijk1966coding}
J.~Schalkwijk, ``A coding scheme for additive noise channels with feedback--{II}: Band-limited signals,'' \emph{IEEE Transactions on Information Theory}, vol.~12, no.~2, pp. 183--189, 1966.

\bibitem{kim2020deepcode}
H.~Kim, Y.~Jiang, S.~Kannan, S.~Oh, and P.~Viswanath, ``Deepcode: Feedback codes via deep learning,'' \emph{IEEE Journal on Selected Areas in Information Theory}, vol.~1, no.~1, pp. 194--206, 2020.

\bibitem{shao2024theory}
Y.~Shao, Q.~Cao, and D.~G{\"u}nd{\"u}z, ``A theory of semantic communication,'' \emph{IEEE Trans. Mobile Compu.}, vol.~23, no.~12, pp. 12\,211--12\,228, 2024.

\bibitem{polyanskiy2010channel}
Y.~Polyanskiy, H.~V. Poor, and S.~Verd{\'u}, ``Channel coding rate in the finite blocklength regime,'' \emph{IEEE Transactions on Information Theory}, vol.~56, no.~5, pp. 2307--2359, 2010.

\bibitem{lau2005proportional}
V.~K. Lau, ``Proportional fair space-time scheduling for wireless communications,'' \emph{IEEE Trans. Comm.}, vol.~53, no.~8, pp. 1353--1360, 2005.

\end{thebibliography}

\end{document}